\documentstyle[11pt,aaspp4]{article}
\journalid{}{}
\articleid{}{}
\slugcomment{}

\begin{document}

\title{The Biconical Outflow in the Seyfert Galaxy NGC~2992}

\author{S. Veilleux\altaffilmark{1,2,3}, P. L.
Shopbell\altaffilmark{1,3,4}, and S. T. Miller\altaffilmark{1}}

\altaffiltext{1}{Department of Astronomy, University of Maryland, College Park,
MD 20742; E-mail: veilleux@astro.umd.edu}

\altaffiltext{2}{Cottrell Scholar of Research Corporation}

\altaffiltext{3}{Visiting Astronomers, Anglo-Australian Telescope}

\altaffiltext{4}{Current Address: Department of Astronomy, MC 105-24, 
California Institute of Technology, Pasadena, CA 91125; E-mail:
pls@astro.caltech.edu}

\begin{abstract}
We report on a detailed kinematic study of the galactic-scale outflow
in the Seyfert galaxy NGC 2992.  The TAURUS-2 Imaging Fabry-Perot
Interferometer was used on the Anglo-Australian 3.9-m telescope to
derive the two-dimensional velocity field of the H$\alpha$-emitting
gas over the central arcminute of NGC 2992. The complete
two-dimensional coverage of the data combined with simple kinematic
models of rotating axisymmetric disks allows us to differentiate the
outflowing material from the line-emitting material associated with
the galactic disk.  The kinematics of the disk component out to R =
3.0 kpc are well modeled by pure circular rotation in a plane inclined
at $i = 68\arcdeg \pm 3\arcdeg$ from the plane of the sky and with
kinematic major axis along P.A. = 32$\arcdeg \pm 3\arcdeg$. The
outflow component is distributed into two wide cones with opening
angle $\approx$ 125 -- 135$\arcdeg$ and extending $\sim$ 2.8 kpc
(18$\arcsec$) on both sides of the nucleus at nearly right angles
($\phi \approx 116\arcdeg \pm 5\arcdeg$) to the disk kinematic major
axis.  The outflow on the SE side of the nucleus is made of two
distinct kinematic components interpreted as the front and back walls
of a cone. The azimuthal velocity gradient in the back-wall component
reflects residual rotational motion which indicates either that the
outflowing material was lifted from the disk or that the underlying
galactic disk is contributing slightly to this component.  A single
outflow component is detected in the NW cone.  A biconical outflow
model with velocities ranging from 50 to 200 km~s$^{-1}$ and oriented
nearly perpendicular to the galactic disk can explain the data.  The
broad line profiles and asymmetries in the velocity fields suggest
that some of the entrained line-emitting material may lie inside the
biconical structure rather than only on the surface of the bicone. The
mass involved in this outflow is of order $\sim$ 1 $\times$ 10$^7$
n$_{e,2}^{-1}$ M$_\odot$, and the bulk and ``turbulent'' kinematic
energies are $\sim$ 6 $\times$ 10$^{53}$ n$_{e,2}^{-1}$ ergs and
$\sim$ 3 $\times$ 10$^{54}$ n$_{e,2}^{-1}$ ergs, respectively. The
most likely energy source is a hot bipolar thermal wind powered on
sub-kpc scale by the AGN and diverted along the galaxy minor axis by
the pressure gradient of the ISM in the host galaxy.  The data are not
consistent with a starburst-driven wind or a collimated outflow powered
by radio jets.
\end{abstract}


\section{Introduction}

Active galactic nuclei (AGN) and nuclear starbursts may severely
disrupt the gas phase of galaxies through deposition of a large amount
of mechanical energy in the centers of galaxies.  As a result, a
large-scale galactic wind (``superwind'') that encompasses much of the
central regions of these galaxies may be created (e.g., Chevalier \&
Clegg 1985; Schiano 1985; Tenorio-Tagle \& Bodenheimer 1988; Tomisaka
\& Ikeuchi 1988; MacLow \& McCray 1988; MacLow, McCray, \& Norman
1989).  Depending upon the extent of the halo and its density and upon
the wind's mechanical luminosity and duration, the wind may ultimately
blow out through the halo and into the intergalactic medium.  The
effects of these superwinds may be far-reaching. Bregman (1978) has
suggested that the Hubble sequence can be understood in terms of a
galaxy's greater ability to sustain winds with increasing
bulge-to-disk ratio. Superwinds may affect the thermal and chemical
evolution of galaxies by depositing large quantities of hot,
metal-enriched material on the outskirts of galaxies. This widespread
circulation of matter and energy between the disks and halos of
galaxies may be responsible for the metallicity-radius relation within
galaxies and the mass-metallicity relation between galaxies (e.g.,
Larson \& Dinerstein 1975; Vader 1986; Franx \& Illingworth 1990;
Carollo \& Danziger 1994; Zaritsky, Kennicutt, \& Huchra 1994;
Jablonka, Martin, \& Arimoto 1996; J{\o}rgensen, Franx, \& Kjaergaard
1996; Pahre, Djorgovski, \& de Carvalho 1998; Trager et al. 1998).
Superwinds also offer a natural way to create a cosmically evolving
population of large, metal-enriched, kinematically-complex gaseous
halos, in many ways resembling the sharp metal lines and Lyman-limit
systems observed in quasar spectra.

Strong evidence for spatially-resolved starburst-driven winds now
exits in several galaxies (e.g., Bland \& Tully 1988; Heckman et
al. 1990; Lehnert \& Heckman 1996; Marlowe et al. 1995; Meurer et
al. 1992; Veilleux et al. 1994).  The case for spatially resolved
outflows in AGN is equally strong (e.g., Cecil, Bland, \& Tully 1990;
Cecil, Wilson, \& Tully 1992; Baum et al. 1993; Wilson et al. 1992;
Veilleux, Tully, \& Bland-Hawthorn 1993; Colbert et al. 1996a, 1996b,
1998) although here the exact mechanism responsible for the outflow is
sometimes uncertain. Possible mechanisms include : (1) thermal wind
from a circumnuclear starburst (e.g., Baum et al. 1993), (2)
collimated radio jets (e.g., Wilson 1981; Taylor, Dyson, \&
Axon 1992), and (3) thermal wind produced by the AGN torus (e.g.,
Krolik \& Begelman 1986; Balsara \& Krolik 1993) or entrained along
the radio ejecta (e.g., Bicknell et al. 1998).

Over the past ten years, our group has been conducting an optical
survey of nearby active and starburst galaxies combining Fabry-Perot
imaging spectrophotometry with radio and X-ray data to track the
energy flow of galactic winds through the various gas phases (see
reviews of recent results in Cecil 2000 and Veilleux 2000). The
complete spatial and kinematic sampling of the Fabry-Perot data is
ideally suited to study the complex and extended morphology of the
warm line-emitting material which is associated with the wind
flow. The radio and X-ray data complement the Fabry-Perot data by
probing the relativistic and hot gas components, respectively.  The
high level of sophistication of recent hydrodynamical simulations
(e.g., Tomisaka 1990; Slavin \& Cox 1992; Mineshige et al. 1993;
Suchkov et al. 1994; Strickland \& Stevens 2000) has provided the
theoretical basis to interpret our data and to predict the evolution
and eventual resting place (disk, halo, or intergalactic medium) of
the outflowing material. In this paper, we present our results on the
Seyfert galaxy NGC 2992.

The proximity of NGC~2992 ($z$ = 0.0078 or 31 Mpc assuming H$_0$ = 75
km s$^{-1}$ Mpc$^{-1}$; see \S 4.2 below) provides a spatial scale
(150 pc arcsec$^{-1}$) which permits a detailed study of the galactic
outflow in this galaxy. The AGN host is a Sa galaxy whose high
inclination ($i$ $\approx$ 70$\arcdeg$) is well suited for studies of
the extraplanar emission. The galactic disk is rich in HI (Hutchmeier
1982) and CO (Sanders \& Mirabel 1985) and is crossed by a disturbed
dust lane oriented along P.A. $\approx$ 30\arcdeg.  Less than
3$\arcmin$ to the south-east of NGC~2992 is the companion galaxy
NGC~2993. The NGC~2992 -- NGC~2993 system (= Arp 245; Arp 1966) is
connected by a HI tidal bridge and displays two spectacular tidal
tails to the north of NGC~2992 and to the south-east of NGC~2993 (see,
e.g., Plate 2 in Ward et al. 1980). The tidal forces associated with
this galactic encounter have long been suspected to be the triggering
mechanism of the nuclear activity in NGC~2992 (e.g., Burbidge et
al. 1972; Osmer, Smith, \& Weedman 1974; Ward et al. 1980).

The evidence for AGN activity in NGC~2992 is seen at nearly all
wavelengths. The pioneering optical work of Burbidge et al. (1972) and
Osmer et al. (1974) revealed a nuclear spectrum dominated by strong
high-excitation emission lines. The detection of a broad (5,500 km
s$^{-1}$) component in the profile of nuclear H$\alpha$ and the
identification of NGC~2992 with a strong HEAO-I X-ray source led Ward
et al. (1978, 1980) to suggest the presence of a partially obscured
AGN in this galaxy. The existence of a broad-line region (BLR) in this
object has since been confirmed at optical wavelengths (e.g., Shuder
1980; V\'eron et al. 1980; Durret \& Bergeron 1988)
and in the near-infrared (e.g., Rix et al. 1990; Goodrich, Veilleux,
\& Hill 1994; Veilleux, Goodrich, \& Hill 1997). X-ray and HI
absorption studies have confirmed the presence of a significant
amount of absorbing material in front of the nucleus of NGC 2992
(N$_{\rm H}$ $\approx$ 10$^{22}$ cm$^{-2}$; Gallimore et al. 1999;
Weaver et al. 1996 and references therein).

The presence of a galactic-scale outflow in NGC~2992 has been
suspected for several years. The 20-cm radio emission from NGC~2992
extends approximately 25$\arcsec$ (3.75 kpc) along the major axis of
the galaxy, but also presents a one-sided 90$\arcsec$ (13.5 kpc)
extension along P.A. $\sim$ 100$\arcdeg$ -- 130$\arcdeg$, i.e. close
to the galaxy minor axis (Ward et al. 1980; Hummel et al. 1983). On
smaller scales, a striking ``figure-8'' structure suggestive of
limb-brightened bubbles or magnetic arches is seen at 6 cm extending
along P.A. = 160$\arcdeg$, about 3$\arcsec$ (5$\arcsec$) north (south)
from a weak nuclear source (Ulvestad \& Wilson 1984; Wehrle \& Morris
1988). The nuclear source has a size $\la$ 0$\farcs$1 -- 0$\farcs$3
(15 -- 45 pc) and appears to have a flat radio spectrum (Condon et
al. 1982; Ulvestad \& Wilson 1984; Unger et al. 1986; Sadler et
al. 1995). A recent re-analysis of the ROSAT HRI data on NGC~2992
(Colbert et al. 1998) reveals that the soft (0.2 -- 2.4 keV) X-ray
emission from this object is extended on a scale of 35 -- 45$\arcsec$
(5.3 -- 6.8 kpc) along the galaxy minor axis. The X-ray nebula is
therefore roughly cospatial with the large-scale radio emission in
this galaxy, suggesting that both are produced by a large-scale
galactic outflow. Supporting evidence for a galactic outflow also
exists at optical wavelengths. The [O~III]-emitting gas is distributed
into two conical regions roughly aligned along the galaxy minor axis
(e.g., Durret \& Bergeron 1987; Werhle \& Morris 1988; Allen et
al. 1999).  Long-slit optical spectroscopy (e.g., Heckman et al. 1981;
Colina et al. 1987; M\'arquez et al. 1998; Allen 1998) indicates that
some of the line-emitting gas in NGC~2992 does not follow simple
galactic rotation. Near-infrared, extended emission embedded within the
northern radio loop was recently detected by Chapman et al. (2000) and
deduced to be of non-stellar origin, adding support to the galactic
outflow scenario.

However, several fundamental questions remain: What is the geometry of
the optical outflow?  How is it oriented with respect to the plane of
the host galaxy?  What are the energetics involved in the outflow?
What is powering the outflow?  The limited spatial coverage of the
long-slit data published so far cannot satisfactorily answer these
questions. In this paper, we present the first complete
two-dimensional velocity field of the line-emitting gas in
NGC~2992. We choose to study the H$\alpha$ emission rather than
[O~III] to reduce the effects of internal dust extinction (important
in this edge-on disk galaxy) and to constrain the kinematics of {\em
both} the galactic disk (bright in H$\alpha$ but faint in [O~III]) and
the outflowing material. Our paper is organized as follows. In \S 2,
we briefly describe the methods used to obtain and reduce our
Fabry-Perot observations.  The technique used to decompose the
emission-line profiles of NGC~2992 into disk and outflow components is
discussed in \S 3. The results of this analysis are discussed in \S 4.
In that section, we compare our Fabry-Perot data with published radio
and X-ray maps, and constrain the three-dimensional kinematics of the
outflow and galactic disk.  In \S 5, we discuss the implications of
our results on the nature and origin of the galactic wind in this
active galaxy. We summarize our results in \S 6 and discuss future
avenues of research.

\section{Observations and Data Reduction}

The Taurus-2 Fabry-Perot Interferometer (Taylor \& Atherton 1980) was
used at the f/8 Cassegrain focus of the 3.9-m Anglo-Australian
Telescope to obtain in dark and photometric conditions 36 five-minute
exposures of redshifted H$\alpha$ in NGC 2992.  These data were
obtained on 1994 March 10 using the Tek 1024 x 1024 detector (read
noise, $\sigma_{\rm R}$ = 2.3 e$^-$, and pixel scale,
0$\farcs$594~px$^{-1}$) under $\sim$ 1$\farcs$5 seeing. A finesse 50
etalon with free spectral range 57 \AA~produced a velocity resolution
of $\sim$ 50 km s$^{-1}$ (this etalon was borrowed from the HIFI
system; Bland \& Tully 1989).  An order separating filter with a
flat-topped transmission profile centered at 6614 \AA~(55 \AA~FWHM)
passed only one etalon order so the true emission line profile of
H$\alpha$ could be synthesized. Exposures were made of diffuse Neon
and flat continuum sources to wavelength calibrate and flatfield the
images, respectively. Finally, the photometric standard star $\eta$
Hya was observed through the same setup as the object to provide an
absolute flux calibration.

These data were reduced using the methods described in Bland \& Tully
(1989), Veilleux et al. (1994), and Bland-Hawthorn (1995).  The
reduction procedure includes flat fielding, spatial registration of
the frames, cosmic ray removal, correction for air mass variations,
sky continuum and line subtraction, correction for the curvature of
the isovelocity surfaces (``phase calibration''), and flux
calibration.  The velocity calibration was verified using the OH sky
line at 6604.13 \AA~(Osterbrock \& Martel 1992).

\section{Data Analysis}

The data reduction produced a spectral datacube with two spatial and
one spectral axes, the latter calibrated in units of wavelength. These
data were spectrally smoothed using a 1/4 -- 1/2 -- 1/4 spectral
filter (Hanning smoothing) and then fit with Gaussian emission-line
profiles in a semi-autonomous fashion, using software developed by one
of us (PLS).  No attempt was made to correct for possible underlying
Balmer absorption. The profile of this feature is dominated by
pressure broadening in the atmospheres of early-type stars, making
this feature indistinguishable from a slowly varying continuum over
the narrow spectral bandpass of our instrument. For the same reason,
the broad H$\alpha$ emission in the nucleus of NGC~2992 was not
detected in our data.

As a first coarse attempt, the H$\alpha$ line in each spectrum was fit
automatically with a single Gaussian, regardless of the profile's
precise shape. These fits allowed us to develop a rough understanding
of both the kinematics and the morphology of the ionized gas
emission. The rotation of the galaxy was clearly evident in the
velocity structure of these fits, as were regions of radially
expanding emission along the galaxy's minor axis. In terms of
morphology, we were able to determine the spatial limits of the
emission region, which eventually encompassed some 2,644 spectra. We
also identified a narrow region through the center of the galaxy where
the prominent dust lane substantially absorbs the H$\alpha$ line
emission. This region, including some 845 spectra, was not processed
further, as the absorbed line profiles are too complex to allow
meaningful Gaussian fits with more than a single velocity component.
This region has been noted on Figure 2. 

Following these coarse fits, a more detailed series of Gaussian fits
were computed, using either one or two distinct velocity components
for each spectrum. In certain cases, a third velocity component could
perhaps be visually discerned, but the difficulty in constraining the
larger number of degrees of freedom in such fits led us to limit
ourselves to two components. In many regions, the presence of split
(i.e., double) line profiles is very evident, and such regions were
manually fit with a pair of Gaussian components, with no constraints
on relative velocity or line width. These manual fits were then
propagated in an automatic fashion across the emission region. Manual
intervention was required where regions of single- and
double-component fits merged. In a few instances, the relative line
widths of the two components were held constant, to provide more
stable convergence of the fits. The result is a collection of 1- and
2-component fits to the H$\alpha$ emission line across the galaxy,
where the line velocities and fluxes are relatively smooth functions
of spatial position. Examples of fits to the line profiles are shown
in Figure 1. The results from these fits are presented in Figures 2, 4 and
5, and are used in the subsequent analysis.

\section{Empirical Results}

\subsection{Distribution of the Ionized Gas}

The distribution of the H$\alpha$-emitting gas in the disk and outflow
components is shown in Figure 2 along with the continuum emission near
6560 \AA~(rest wavelength). These flux maps were corrected for
extinction from our own Galaxy [E(B--V) = 0.060 mag; Burstein \&
Heiles 1982; Schlegel, Finkbeiner, \& Davis 1998]
but not for intrinsic reddening (these corrections will be carried out
in \S 5.1). The well-known dust lane is easily visible along
P.A. $\approx$ 30$\arcdeg$ in the continuum image.  The total
H$\alpha$ flux in these maps amounts to 1.6 $\times$ 10$^{-12}$
erg$^{-1}$ cm$^{-2}$ s$^{-1}$.
Our combined H$\alpha$ flux map (Fig. 2$b$) is very similar to the
H$\alpha$ + [N~II] $\lambda$6583 narrow-band images of Wehrle \&
Morris (1988) and Allen et al. (1999).
All of the features seen in their images within $\sim$ 1$\arcmin$ of
the nucleus are also detected in our H$\alpha$ map. 
Note the absence of a disk component in the SE quadrant; the emission
from the outflowing material in this region overwhelms the emission
from the disk (see \S\S 4.2 and 4.3).

The outflow component (Fig. 2$d$) is distributed into two wide cones
which extend up to $\sim$ 18$\arcsec$ (2.8 kpc) from the peak in the
optical continuum map.  Both cones have similar opening angles of
order 125$\arcdeg$ -- 135$\arcdeg$.  The bisectors of each of the
cones coincide with each other and lie along P.A. $\approx$
116$\arcdeg$, or almost exactly perpendicular to the kinematic major
axis of the inner disk (P.A. $\approx$ 32$\arcdeg$). This strongly
suggests that the axis of the bicone is perpendicular to the galactic
disk, and that the material in the SE cone is emerging from under the
galaxy disk while the material in the NW cone is emerging from above
the disk. If this is the case, the north-west rim of the disk has to
be on the near side (see also discussion in Allen et al. 1999).

Figure 3 compares the X-ray (0.2 -- 2.4 keV) and radio (6-cm) emission
from NGC~2992 with the optical line emission from the outflow
component. The RA/DEC coordinates of the Fabry-Perot data were
determined by comparing the pixel positions of 5 stars in the field
with the corresponding field in the (2nd generation) Digitized Sky
Survey. Given the pixel scales of our data and the DSS, as well as the
relatively small number of available comparison stars, we estimate the
positional accuracy of this method to be $\la$ 1$\arcsec$. The
absolute positions of the radio maps should be more accurate than
this.  The X-ray map shown Figure 3 is the ROSAT/HRI image originally
published by Weaver et al. (1996) and reanalyzed by Colbert et
al. (1998). This image was smoothed with a Gaussian kernel of $\sigma$
= 24$\arcsec$ to bring up the faint X-ray bridge which connects
NGC~2992 to its companion NGC~2993 and to show the faint extended
emission along P.A. $\approx$ 112$\arcdeg$ (Colbert et al. 1998). This
faint extension lies close to the axis of the outflow bicone,
suggesting a possible link between the hot and warm gas phases in the
outflow. The radio map shown in the upper right panel of Figure 3 is
based on the VLA C-configuration data of Colbert et al. (1996a) using
3$\arcsec$ uniform weighting (not shown in the original paper). The
high-resolution radio data (lower right panel in Fig. 3) were taken by
Ulvestad \& Wilson (1984) using the VLA A-configuration with a spatial
resolution $\sim$ 0$\farcs$4.  The SE outflow cone overlaps with the
southern portion of the ``figure-8'' structure seen in the radio. The
base of the NW outflow cone lies close to the outer edge of the
northern radio ``loop''.  The position angles of the outflow bisector
and the long axis of the ``figure-8'' structure differ by $\sim$
25$\arcdeg$ -- 35$\arcdeg$, but the one-sided 13.5-kpc radio extension
detected by Ward et al. (1980) and Hummel et al. (1983) lies roughly
along the same direction as the mid-axis of the SE optical cone
(P.A. $\sim$ 100$\arcdeg$ -- 130$\arcdeg$; arrows mark this direction
in the right panels of Fig. 3). Interestingly, the high-velocity NW
cloud shown in Fig.  4$c$ lies roughly along the major axis of the
inner radio structure.

\subsection{Velocity Field of the Disk Component}

The velocity fields and line width maps of the disk and outflow
components are shown in Figures 4 and 5. The uncertainty on these
velocities and line widths is $\sim$ 20 km s$^{-1}$ in the brighter
disk H~II regions, but may be 2-3 times larger in the fainter
material. The velocity field of the disk component shows the typical
``spider diagram'' produced by differential rotation. None of the gas
in the SE quadrant appears to follow normal galactic rotation; the
kinematics of this gas are therefore presented in Figs. 4$b$ and 4$c$
and discussed in the next section. The average line width in the disk
component is $\sim$ 170 km s$^{-1}$, being significantly larger in the
dust lane where selective dust extinction sometimes produces complex
profiles (Fig. 5$a$).  In Figure 4$a$ a string of black dots traces
the steepest gradient through the observed velocity field (using the
method described in Bland 1986); this is the kinematic line of nodes.
Figure 6 shows the H$\alpha$ rotation curve along this locus. It rises
linearly in the inner region out to a radius of $\sim$ 1 kpc where it
reaches a local maximum of $\sim$ 190 (NE) -- 200 (SW) $\pm$ 20 km
s$^{-1}$ (or a deprojected value of $\sim$ 205 (NE) -- 215 (SW) $\pm
20$ km s$^{-1}$ if $i$ = 68$\arcdeg$; see below). Beyond this radius,
the amplitude of the rotation curve first decreases with increasing
radius and then rises again to reach a deprojected velocity amplitude
of $\sim$ 225 $\pm$ 20 km s$^{-1}$ and perhaps a value even larger at
R $\ga$ 3 kpc. A systemic velocity of $\sim$ 2335 $\pm$ 20 km s$^{-1}$
is derived from this rotation curve.  This value is confirmed in our
more detailed two-dimensional analysis (see below). This value is in
good agreement with estimates from other optical datasets [e.g.,
Nelson \& Whittle (1995) and M\`arquez et al. (1998) both derived a
systemic velocity of 2330 km s$^{-1}$ using stars and the
line-emitting gas, respectively; see also Keel 1996], but slightly
exceeds the value derived from HI data (e.g., 2314 km s$^{-1}$ from de
Vaucouleurs et al. 1991). This apparent discrepancy is probably due to
a slight asymmetry between the inner optical rotation curve ($R \la $
3 kpc) and the outer H~I rotation curve.  Tidal forces from the
companion galaxy NGC~2993 are likely to disturb the velocity field of
the HI material more severely than that of the inner disk.

To better quantify the velocity field of the disk component, we
compared our data with simple kinematic models which approximate the
intrinsic gas orbits as nonintersecting, inclined elliptical annuli
that conserve angular momentum (see Bland 1986 or Staveley-Smith et
al. 1990 for a detailed discussion of these models).  We began our
analysis by exploring the parameter space for an axisymmetric disk
with inclination 60$\arcdeg$ $\le$ $i$ $\le$ 80$\arcdeg$, kinematic
major axis 0$\arcdeg$ $\le$ P.A. $\le$ 45$\arcdeg$, and systemic
velocity 2,250 $\le$ $V_{\rm sys}$ $\le$ 2,400 km s$^{-1}$. This
region of parameter space was selected based on the results of
previous optical and HI kinematic studies of NGC~2992.  We tried to
use either (a) a smoothed and symmetric version of the rotation curve
shown in Figure 6 or (b) one of Brandt's (1960) rotation curves.  In
the first case, the NE and SW sections of the galaxy rotation curve
were averaged together and then fit with a fourth-order polynomial.
When searching for the best-fitting model, only the inner portion of
the disk out to $\sim$ 3 kpc was considered because the velocity field
beyond this point is incomplete and possibly affected by the tidal
interaction with NGC 2993.

Figure 7$a$ shows the best-fitting axisymmetric model ($i$ = 68$\arcdeg$,
P.A. = 32$\arcdeg$, and $V_{\rm sys}$ = 2,335 km s$^{-1}$). 
The smoothed and symmetric version of the rotation curve was found to
give a better fit to the data than the Brandt models. Figure 7$b$
shows the residuals after subtracting the best-fitting model from our
measured H$\alpha$ velocity field.  The residuals are centered on 0 km
s$^{-1}$ with a standard deviation of 32 km s$^{-1}$. Overall, this
axisymmetric model is a good match to the data. We find no evidence
for ``twisting'' of the isovelocity contours in the disk component
which would signal the presence of a bar potential.  The largest
positive residuals between the data and the model are seen $\sim$
10$\arcsec$ -- 20$\arcsec$ from the nucleus along P.A. $\approx$
40$\arcdeg$. Negative discrepancies are detected in the SW portion of
the disk. These discrepancies are due in most part to our use of a
symmetric rotation curve in the model; slight asymmetries between the
NE and SW sides of the disk are visible in Figures 4$a$ and 6.

Excellent agreement is found between our disk velocity field and the
published long-slit data of M\`arquez et al. (1998).  The model of
M\`arquez et al. (1998) is similar to our best-fitting disk model:
their values of the kinematic major axis (30$\arcdeg$), inclination
(70$\arcdeg$), velocity amplitude (250 km s$^{-1}$), and systemic
velocity (2,330 km s$^{-1}$) all lie within the uncertainties of the
measurements. However, the two-dimensional coverage of our data allows
us to constrain the shape of the rotation curve slightly better,
emphasizing in particular the presence of a peak in the velocity curve
at $\sim$ 1 kpc on both sides of the nucleus. The velocities of Colina et
al. (1987) also agree with ours, although here the comparison is more
difficult because their measurements do not extend as far from the
nucleus as those of M\`arquez et al.

\subsection{Velocity Field of the Outflow Component}

Two distinct kinematic components are detected in the SE cone
(Fig. 4).  None of these components shares the kinematics expected
from simple galactic rotation (Fig. 7$a$). One of the components is
significantly blueshifted with respect to the expected rotation field
of the disk while the other is significantly redshifted. The
redshifted component also presents a azimuthal velocity gradient
indicative of residual rotational motion, which suggests that the
outflowing material was lifted from the disk or that the underlying
galactic disk is contributing slightly to this component.  No
significant gradient is detected in the blueshifted component.  The
widths (FWHM) of both components range from 50 km~s$^{-1}$ to more
than 350 km~s$^{-1}$, with flux-weighted averages of 200 and 150 km
s$^{-1}$ for the blueshifted and redshifted components, respectively
(Fig. 5).

With the exception of the high-velocity (170 -- 250 km $^{-1}$
relative to the systemic velocity) NW cloud, more evident in Figures 4
and 5, the outflow in the NW cone is well modeled by a single
emission-line component with FWHM ranging from 75 km~s$^{-1}$ to about
350 km~s$^{-1}$ (flux-weighted average of 200 km~s$^{-1}$).  That is
to say most of the substructures detected in the emission-line
profiles from the NW quadrant of NGC~2992 are due to the kinematic
superposition of the disk and outflow components rather than to the
presence of substructure in the outflow itself.

This kinematic asymmetry between the NW and SE outflows is somewhat
puzzling.  In most wide-angle starburst- and AGN-driven outflows (see
results of surveys by Heckman et al. 1990; Lehnert \& Heckman 1996;
Colbert et al. 1998; Veilleux 2000), double-peaked profiles are
detected on both sides of the nucleus, implying that the optical
line-emitting gas lies on the surface of expanding bubbles or a
biconic structure. The implied velocities in these powerful outflows
typically are of order a few hundred km s$^{-1}$ (although exceptions
exist: e.g., the outflow velocities in NGC~3079 exceed 1,500 km
s$^{-1}$; Veilleux et al. 1994). The absence of line splitting on one
side is sometimes due to obscuration by the disk (e.g., NGC~3079). Can
the same thing be happening in NGC~2992? The geometry of the biconical
outflow in NGC~2992 (shown in Figure 8) is such that the SE outflow
cone is in front of the galaxy disk and the base of the NW outflow is
behind it.  This may explain the large strength of the SE outflow
components relative to the (undetected?) disk component. However, if
the gas flows along the walls of the cones, one would expect to detect
the blueshifted component of the NW outflow (which is located on the
near side of the cone) more easily than the redshifted component; the
opposite is seen in NGC~2992. High-resolution HST images of the
outflows in M82 (Shopbell et al. 2000) and NGC~3079 (Cecil et
al. 2000) indicate that the entrained material is not distributed
uniformly on the surface of the bicones (M82) or bubble (NGC~3079),
but instead forms a complex network of clouds and filaments where some
regions are sometimes devoid of line-emitting material.  The near side
of the NW cone of NGC~2992 may just be such a region. The absence of warm
(T $\approx$ 10$^4$ K) line-emitting gas in this region does not
preclude the existence of hot or molecular material which would only be
detectable in the X-rays or at millimetric wavelengths, respectively.

The biconical outflow model may be constrained further by comparing
the velocities of the various kinematic components of the outflow.
Using the nomenclature and geometry of Figure 8, we have $\vert V_{\rm
blue}{\rm (SE)}\vert$ = $V_1$~cos $\alpha$, $\vert V_{\rm red}{\rm
(NW)}\vert$ = $V_4$~cos $\alpha$ and $\vert V_{\rm red}{\rm
(SE)}\vert$= $V_2$~cos $\beta$ where $\alpha = i - \theta/2 \approx
3\arcdeg$ (assuming $\theta \approx 130\arcdeg$), $\beta = 180\arcdeg
- \theta - \alpha = 47\arcdeg$, and $V_1$ (= $V_4$ if the outflow is
bi-symmetric) and $V_2$ are the outflow velocities at the top and at
the base of the bicone, respectively (see Fig. 8). From our data, we
measure $\vert V_{\rm red}{\rm (NW)}\vert$ = 25 -- 220 (100)
km~s$^{-1}$, $\vert V_{\rm blue}{\rm (SE)}\vert$ = 15 -- 140 (40)
km~s$^{-1}$, and $\vert V_{\rm red}{\rm (SE)}\vert$ = 25 -- 100 (40)
km~s$^{-1}$, where the values in parentheses are the flux-weighted
($\approx$ mass-weighted) averages.  For this calculation, the
redshifted velocities in the SE cone were corrected for residual
rotational motion by subtracting the expected disk velocity field
(Fig. 7$a$) from our measured velocities.  The results are not
consistent with a constant outflow velocity: $V_1$ = 25 -- 100 (40)
km~s$^{-1}$, $V_2$ = 40 -- 150 (60) km~s$^{-1}$, and $V_4$ = 25 -- 220
(100) km~s$^{-1}$. Possible trends of decreasing outflow velocities
with increasing distance from the nucleus may be present in the $V_2$
and $V_4$ components.

Given the complexity of the velocity field in the outflow component,
we do not feel that more elaborate modeling of the data is warranted
at this stage.  Hence, there is no simple way of testing our
assumption that the gas flows along the surface of the
bicone. However, the large line widths of each of the three main
outflow components suggest that the outflow is highly ``turbulent''
and that each of these components may in fact be the superposition of
several filaments at different depths along our line sight. The broad
range of velocities and general lack of line splitting in the NW cone
may indicate that the line-emitting gas lies inside the biconical
structure, filling in some of the velocity space between the front and
back walls of the outflow. In this type of mass-loaded outflow, the
entrained gas is mixed in with the wind material. In this respect, the
NW wind of NGC~2992 may share a strong resemblance with the AGN-driven
winds of Circinus (e.g., Veilleux \& Bland-Hawthorn 1997; Elmouttie et
al. 1998) and NGC~4388 (e.g., Veilleux et al. 1999).

\section{Discussion}

\subsection{Dynamical Timescale, Masses, and Energies}

The dynamical time scale of the outflow can be estimated from the
linear extent of the outflow and an estimate of the average velocity:
\begin{equation}
t_{\rm dyn} \approx R/V = 9.8 \times 10^6~(R_{\rm obs}/2~{\rm
kpc})~(V_{\rm obs}/200~{\rm km~s}^{-1})^{-1}~~~{\rm yr}
\end{equation}
Projection effects may affect this timescale by a factor of $\sim$
2. 

The ionized mass taking part in the outflow follows from the H$\alpha$
flux.  In the following discussion, we parametrize the mass in terms
of the density and adopt an electron temperature of 10$^4$ K (see,
e.g., Allen et al. 1999 for measurements of the temperature and
density of the ionized material in NGC~2992).  The total H$\alpha$
intensity (luminosity) of $\sim$ 4.3 $\times$ 10$^{-13}$ erg s$^{-1}$
cm$^{-2}$ (4.9 $\times$ 10$^{40}$ ergs s$^{-1}$) from the material
involved in the outflow yields an ionized mass of $\sim$ 2 $\times$
10$^6$ n$_{e,2}^{-1}$ M$_\odot$, where n$_{e,2}$ is the electron
density in units of 100 cm$^{-3}$. In this calculation, we assumed
Case B recombination and used an effective recombination coefficient
for H$\alpha$ of 8.6 $\times$ 10$^{-14}$ cm$^{-3}$ s$^{-1}$
(Osterbrock 1989). This number has to be corrected for dust
obscuration from our Galaxy and intrinsic to NGC~2992.  Here, we adopt
the average reddening values in the NW and SE cones derived by Allen
et al. (1999): $A_V$ = 0.86 and 1.8, respectively. Using these values,
we get
\begin{equation}
M^{\rm corr} \approx 1 \times 10^7 n_{e,2}^{-1}~~{\rm M}_\odot. 
\end{equation}
The rate of mass outflow can be estimated if it has been constant over
the dynamical age of the outflow:
\begin{equation}
dM/dt \sim M^{\rm corr}/t_{\rm dyn} \approx~n_{e,2}^{-1}~~{\rm M}_\odot~{\rm yr}^{-1}.
\end{equation}
Given the low velocity of the outflow relative to the escape velocity
of NGC~2992 (V$_{\rm esc}(r)$ $>$ $\sqrt2$~V$_{\rm rot}(r)$ $\approx$
300 km~s$^{-1}$), the ionized material is likely to be deposited in
the halo of the galaxy. This mass outflow rate is a lower limit
because it includes only the contribution from the warm ionized
H$\alpha$-emitting material (see \S 5.2).

To first order, the bulk kinetic energy of the outflowing gas, $E_{\rm
bulk}$, may be derived by summing the bulk kinetic energy over each
pixel in the outflow components ($\Sigma_i$ 1/2 $\delta m_i$ $v_i^2$,
where $\delta m_i$ and $v_i$ are the mass and observed radial velocity
at pixel $i$). Here, we use the rotation-corrected velocity field for
the redshifted component in the SE cone while the velocities in the
other components are measured with respect to the systemic velocity,
2335 km~s$^{-1}$. We find $E_{\rm bulk}$ $\approx$ 2 $\times$
10$^{53}$ $f_{\rm proj}$~n$_{e,2}^{-1}$ ergs where $f_{\rm proj}$
$\approx$ 1 -- 3 is the correction factor which accounts for
projection effects. The ``turbulent'' kinematic energy of the outflow
component, $E_{\rm turb}$, may be derived from $\Sigma_i$ $\delta m_i$
$\sigma_i^2$, where $\sigma_i$ = FWHM/1.67 and FWHM is the full width
at half maximum of the line profiles corrected in quadrature for the
finite instrumental profile (FWHM$_i$ = 50 km s$^{-1}$); we get
$E_{\rm turb}$ $\approx$ 7 $\times$ 10$^{53}$ n$_{e,2}^{-1}$ ergs.
Applying the reddening corrections of Allen et al. (1999) to each
cone, we obtain
\begin{equation}
E^{\rm corr}_{\rm bulk} \approx 6 \times 10^{53} f_{\rm
proj}~n_{e,2}^{-1}~~{\rm ergs}
\end{equation}
and 
\begin{equation}
E^{\rm corr}_{\rm turb} \approx 3 \times 10^{54} n_{e,2}^{-1}~~{\rm ergs.}
\end{equation}
Note that the ``turbulent'' component of the kinetic energy dominates
the outflow in NGC~2992.

Using n$_{e,2}$ = 1 as a representative value of the electron density
of the outflowing material (Allen et al. 1999) and taking into account
projection effects, the total kinetic energy of the outflowing gas is
$E_{\rm kin}$ = $E^{\rm corr}_{\rm bulk}$ + $E^{\rm corr}_{\rm turb}$
$\approx$ 4 $\times$ 10$^{54}$ ergs.  Using our estimate of the
dynamical timescale, we can derive the time-averaged rate of kinetic
energy injected into the circumnuclear region:
\begin{equation}
dE_{\rm kin}/dt \simeq E_{\rm kin}/t_{\rm dyn}~\approx 10^{40}~{\rm n}_{e,2}^{-1}~~~{\rm erg~s}^{-1}.
\end{equation}
The uncertainty on this number is large (perhaps as much as an order
of magnitude) because of unknown projection effects and imperfect
reddening correction.  Moreover, we must emphasize again that the
energetics derived in this section only take into account the optical
ionized component of the outflow.  Entrainment of the molecular and
neutral gas from the disk could significantly increase the energetics
of the outflow (e.g., M82: Wei\ss~et al. 1999). The kinetic energy of
the ionized outflow in NGC~2992 is comparable to that in the
starburst-driven outflow of M82 (Shopbell \& Bland-Hawthorn 1998), the
AGN-driven outflow of NGC~1068 (Cecil et al. 1990), and even the
powerful superwind-blown bubble of NGC~3079 (Veilleux et al. 1994;
Cecil et al. 2000). This is somewhat surprising given the relatively
small outflow velocities observed in NGC~2992. This is a consequence
of the large amount of ionized material entrained in the outflow
(eqn. 2) and of the large ``turbulent'' component to the kinetic
energy.

\subsection{Origin of the Outflow}

In this section, we return to the question of the origin of the outflow in
NGC~2992.  Three scenarios were mentioned in the Introduction: (1)
thermal wind from a circumnuclear starburst, (2) collimated radio jet,
and (2) thermal wind produced by the AGN.

\subsubsection{Starburst-Driven Wind}

This scenario is attractive because it can readily explain the
orientation of the outflow along the galaxy minor axis. In this
picture, the kinetic energy from strong stellar winds of massive young
stars and from type-II supernovae is thermalized, producing a hot,
overpressured cavity which bursts out along the minor axis of the
galaxy disk entraining with it some of the ambient disk material
(e.g., Suchkov et al. 1994, 1996; Strickland \& Stevens 2000). A
fundamental problem with this scenario is the lack of direct evidence
for the presence of a circumnuclear starburst in NGC~2992. The
extensive long-slit spectroscopic study by Allen et al. (1999) shows
no obvious signs for an on-going nuclear starburst at optical
wavelengths (e.g., Wolf-Rayet features).  Near-infrared spectroscopy
of the nuclear region detects relatively strong H$_2$ and [Fe~II]
emission possibly indicative of star formation near the nucleus of
NGC~2992, but this emission can also be produced by shocks associated
with the nuclear outflow (Goodrich et al. 1994; Veilleux et al. 1997).
Substantial amounts of warm dust ({\em IRAS} 25-$\mu$m--to--60-$\mu$m
flux ratio $f_{25}/f_{60} = 0.22$ and far-infrared luminosity L$_{\rm
FIR}$ = 2.9 $\times$ 10$^{10}$ L$_\odot$; Spinoglio \& Malkan 1989)
and neutral and molecular gas are present near the nucleus of NGC~2992
(e.g., Gallimore et al. 1999; Maiolino et al. 1997; Sanders \& Mirabel
1985), but there is no evidence that this gas is forming stars at a
rate sufficient to sustain the outflow. Finally, there is no sign of a
sub-kpc circumnuclear starburst at radio wavelengths. The peculiar
``figure-8'' radio structure detected on that scale (Ulvestad \&
Wilson 1984; Wehrle \& Morris 1988) is probably of synchrotron origin
and is believed to have been produced by the AGN (Wehrle \& Morris
1988).

\subsubsection{Collimated Radio Jet}

The radio morphology is also a problem for the collimated jet-driven
outflow model. NGC~2992 does not display the sub-kpc ``linear'' radio
structures seen on arcsecond scale in $\ga$~30\% of Seyfert galaxies
(e.g., Ulvestad \& Wilson 1989; Pedlar et al. 1993 and references
therein). The resemblance of these linear structures to jet- and
lobe-like structures in radio-loud AGNs has been used to argue for the
existence of low-power nuclear jets in these Seyfert galaxies (e.g.,
Kukula et al. 1999 and references therein). The diffuse radio
structure of NGC~2992 suggests a more quiescent deposition of energy
into the surrounding medium. Equally damaging for the collimated jet
model is the large opening angle of the optical outflow and the
misalignment between the major axis of the ``figure-8'' radio
structure seen on arcsecond scale and the axis of the optical outflow
(Fig. 3). Radio-optical alignments are often seen in Seyferts with
``linear'' radio structure (e.g., Whittle et al. 1988; Haniff, Wilson,
\& Ward 1988) and in powerful radio galaxies (e.g., Chambers, Miley,
\& van Breugel 1987). Alignment between the optical ionization cones
and the nuclear radio source is also frequently seen in Seyfert
galaxies (e.g., Wilson \& Tsvetanov 1994). The absence of bright radio
knots outside of the nucleus of NGC~2992 seems to rule out the
possibility that the radio jet has been deflected by a dense cloud in
the circumnuclear region (as in NGC~1068; Gallimore, Baum, \& O'Dea
1996).

However, the radio data of NGC~2992 do not rule out the possibility
that the ``figure-8'' structure represents radio lobes or bubbles of
relativistic plasma (``plasmons'') fed on small ($\la$~0$\farcs$3
$\approx$ 50 pc) scale by collimated jets.  The linear, supersonic
motions of these plasmons may drive bowshocks into the ambient nuclear
medium. Several authors have studied the effects of a supersonic light
jet or plasmon propagating into the circumnuclear medium of a galaxy
(e.g., Pedlar, Dyson, \& Unger 1985; Taylor et al. 1989; Taylor,
Dyson, \& Axon 1992; Ferruit et al. 1997a; Steffen et al. 1997). Using
classical equipartition arguments, one finds that the pressure in the
relativistic plasma is inadequate to drive high-velocity shocks into
the interstellar medium and produce the observed emission line
flux. In all these models, the cooled shocked gas is therefore
photoionized mainly by the ultraviolet nuclear continuum. These models
have had significant success explaining the ENLR of Seyfert galaxies
with linear radio structures (e.g., Whittle et al. 1986, 1988; Unger
et al. 1987; Taylor et al. 1992; Capetti et al. 1995a,b; Bower,
Wilson, \& Mulchaey 1994; Bower et al. 1995; Ferruit et al. 1997b;
Axon et al. 1998), but to our knowledge it has never been successfully
applied to galaxies with diffuse radio morphology like NGC~2992.  The
lack of spatial correlation between the ridges of radio emission
(e.g., the ``figure-8'' structure) and the optical line-emitting
material participating in the outflow in NGC~2992 is difficult to
reconcile with the predictions of this model.

\subsubsection{Thermal Wind from the AGN}

Recently, Bicknell et al. (1998) have argued that the internal energy
densities of many Seyfert jets are dominated by thermal plasma rather
than relativistic plasma, as it is generally believed to be the case
in the more powerful radio galaxies.  Focussing on five well-studied
Seyferts including NGC~1068, they have shown that an important
fraction of the energy and momentum flux originally carried by the
radio-emitting jet may be contained in entrained thermal plasma, which
is not detected in synchrotron emission. Veilleux et al. (1999)
recently argued for a similar mechanism to explain the outflow in
NGC~4388 (see also Colbert et al. 1996a, 1998).  These ``jet-driven''
thermal winds are fundamentally different from the ``torus-driven''
wind of Krolik \& Begelman (1986; see also Balsara \& Krolik 1993)
because the geometry of the former is regulated by that of the radio
ejecta rather than by the opening angle of the sub-pc torus. However,
these two types of winds may be difficult to differentiate on the kpc
scale of the optical outflow in NGC~2992, given the diffuse radio
morphology of this object on that scale.

Allen et al. (1999) recently applied this jet-driven wind model to the
narrow-line region of NGC~2992 and have argued based on the
emission-line ratios that the ENLR in NGC~2992 is powered by shocks
with $V_s \approx 300 - 500$ km~s$^{-1}$.  Our analysis of the
galactic outflow now provides additional constraints on this model. As
discussed in \S 4.1, we find the axis of the optical outflow bicone to
be roughly aligned with the minor axis of the galaxy and the radio
structure on large scale, but to be misaligned by $\sim$ 44$\arcdeg$
with respect to the major axis of the ``figure-8'' radio structure on
arcsecond scale. As first suggested by Ulvestad \& Wilson (1984), this
misalignment may be caused by the differential thermal pressure across
the radio structure due to static pressure gradients in the
interstellar gas (``buoyancy effects'', Henriksen, Vall\'ee, \& Bridle
1981; Smith \& Norman 1981; Fiedler \& Henriksen 1984).  The net
buoyancy force on the radio ejecta of density $\rho_c$ immersed in a
region of density $\rho_0$ will be $\propto$ $\rho_0 - \rho_c$, and
will act to bend the trajectory of the radio plasma away from the
major axis of the galaxy as seen in NGC~2992. Note that the large
change in direction could not be explained in this fashion for the
collimated-jet model (\S 5.2.2) because only the pressure difference
across the jet thickness can contribute to the bending (Fiedler \&
Henriksen 1984). 

A potentially serious problem with the jet-fed thermal wind model is
the relatively small outflow velocities detected in NGC~2992 (recall
that the flux-weighted average of the apparent outflow velocities is
less than 100 km~s$^{-1}$; \S 4.3). Shock velocities of 300 -- 500
km~s$^{-1}$ are needed to reproduce the line ratios in the ENLR of
NGC~2992 (Allen et al. 1999).  Although projection effects undoubtedly
affect our measurements of the velocities, the corrections are
probably not severe [see Fig. 8; the angle $\alpha$ between our line
of sight and the near-side (far-side) surface of the SE (NW) cone is
only $\sim$ 3$\arcdeg$ so our viewing angle should allow us to measure
most of the actual outflow velocity if the motion is purely radial].
However, one needs to be cautious when comparing shock velocities with
gas outflow velocities: the outflow velocities derived from our
Fabry-Perot data reflect the motion of the (slow moving) dense
entrained material and are almost certainly smaller than the actual
shock velocity.

Evidence for high-velocity shocks in NGC~2992 may be found at other
wavelengths.  Shocks with velocity $V_s \approx 400$ km~s$^{-1}$ will
heat the ISM to a temperature
\begin{equation}
T_s \approx 2.2 \times 10^6~(V_s/400)^2 {\rm ~K} \approx 
193~(V_s/400)^2 {\rm ~eV}.
\end{equation}
This gas may be responsible for some of the extended X-ray emission
detected by Colbert et al. (1998). Rough estimates of the pressure of
the X-ray emitting gas by Colbert et al. (1998) suggest that it is
possibly well above the conventionally calculated non-thermal pressure
-- this hot gas may therefore be a dynamically important component of
the outflow. Given the current estimates of its density (Colbert et
al. 1998), the X-ray emitting gas may also be an important contributor
to the rate of mass (and metal) outflow in this object (see \S
5.1). Definite answers to these questions will have to wait until
Chandra data on this object become available. These data may also be
able to constrain the morphology (jet-like or wide-angled?) of the
X-ray gas on sufficiently small scale to help discriminate between the
jet-driven wind model of Bicknell et al. (1998) and the torus-driven
wind model of Krolik \& Begelman (1986) and Balsara \& Krolik (1993).

The optical line emission will be produced once the gas cools to
$\sim10^4$ K after
\begin{equation}
t_c \approx 3.2 \times 10^3~ n_s^{-1}~T_s^{1.5}~\simeq 3 \times
10^{5} n_s^{-1}~(V_s/400)^3~~~{\rm yr},
\end{equation}
where $n_s$ is the postshock density or four times the ambient proton
density. The line-emitting gas will be displaced from the shock front
by 
\begin{equation}
l_c \approx t_c~V_s~\approx~36~n_0^{-1}~(V_s/400)^4~~~{\rm pc}.
\end{equation}
Following Dopita \& Sutherland (1996), the total energy dissipated in
the shock each second is
\begin{equation}
L_{\rm shock} \simeq {1\over 2} \rho A V_s^3  = 7.9 \times
10^{27}~n_a~A~V_s^3 {\rm ~~erg~s}^{-1},
\end{equation}
where $A$ is the shock surface area in pc$^2$, and the H$\alpha$
luminosities from the shock and the precursor are
\begin{equation}
L(H\alpha)_{\rm shock} \approx 3 L(H\beta)_{\rm shock} = 6.0 \times
10^{33}~n_a~A~(V_s/400)^{2.41} {\rm ~~erg~s}^{-1}
\end{equation}
and
\begin{equation}
L(H\alpha)_{\rm precursor} \approx 3 L(H\beta)_{\rm precursor} = 6.6
\times 10^{33}~n_a~A~(V_s/400)^{2.28} {\rm ~~erg~s}^{-1},
\end{equation}
respectively. Using an ambient (preshock) ISM density of $\sim$1 {\rm
~cm}$^{-3}$ and $A \simeq 2 \times \pi~(1,000)^2 \approx 6 \times
10^6$ pc$^2$ (this is the area at the top of the bicones), we find
that the expected H$\alpha$ luminosity of the shock+precursor model is
similar to the dereddened H$\alpha$ luminosity of the outflow
component.  Considering the large uncertainties on both the observed
and predicted values, it seems physically possible to power a large
fraction of the line emission across the bicone with shocks. Allen et
al. (1999) came to the same conclusion using line ratio arguments.

\section{Conclusions}

The complete two-dimensional coverage of our Fabry-Perot data has
allowed us to decompose the circumnuclear emission-line region of
NGC~2992 into two distinct kinematic components: a rotational
component which lies in the plane of the galactic disk and an
outflowing component which traces a biconical wind along the minor
axis of the galactic disk. Outflow velocities ranging from 50 to 200
km~s$^{-1}$ are detected in the biconical region. The presence of two
distinct kinematic components in one of the cones suggest that some
(but probably not all) of the warm ionized material lies on the
surface of the bicone.  The mass ($\sim$ 1 $\times$ 10$^7$
n$_{e,2}^{-1}$ M$_\odot$) and kinematic energy ($\sim$ 4 $\times$
10$^{54}$ n$_{e,2}^{-1}$ ergs) involved in this outflow are typical of
other starburst- and AGN-driven outflows.  Evidence for a slight
rotational component in some of the entrained material located in
front of the galactic disk suggests that this material has been lifted
from the disk or that emission from the underlying disk is
contributing to this component.  The diffuse radio morphology and
absence of direct evidence for a powerful nuclear starburst in this
galaxy seem to rule out collimated radio jets or starburst-driven wind
as the source of the outflow. By process of elimination, the most
likely scenario is that of a thermally-dominated wind diverted along
the galaxy minor axis by the pressure gradient of the ISM and powered
on sub-kpc scale by (yet undetected) low-energy radio jets. This
picture is consistent with the jet-driven outflow model of Colbert et
al. (1996a), Bicknell et al. (1998) and Allen et al. (1999). A similar
process was suggested by Veilleux et al. (1999) to explain the outflow
in NGC~4388.  In NGC~2992, shock velocities of $\sim$ 400 km~s$^{-1}$
would be sufficient to power the optical line emission from the
entrained warm ionized material.

The currently available data on NGC~2992 do not allow us to
distinguish between the jet-driven wind of Bicknell et al. (1998) and
the torus-driven wind of Krolik \& Begelman (1986). Deep radio maps on
VLBI scale will be useful to look for the presence of a collimated
pc-scale jet. Chandra data of this object would also be helpful to
constrain the geometry, temperature and pressure of the X-ray emitting
gas and hence the dynamical role of the hot thermal gas in the
galactic outflow of NGC~2992. The shock velocities derived from the
X-ray data could be directly compared with those necessary to ionize
the ENLR clouds.

\clearpage

\acknowledgments

We thank J. Bland-Hawthorn for help during the observations and
analysis and R. B. Tully for the loan of the HIFI etalon and
filters. The VLA data used in this analysis were graciously provided
to us by A. S. Wilson and E. J. M. Colbert. SV is grateful for partial
support of this research by a Cottrell Scholarship awarded by the
Research Corporation, NASA/LTSA grant NAG 56547, and NSF/CAREER grant
AST-9874973. This research has made use of the NASA/IPAC Extragalactic
Database (NED), which is operated by the Jet Propulsion Laboratory,
California Institute of Technology, under contract with the National
Aeronautics and Space Administration.

\clearpage

\begin{figure}
\caption{Examples of one- and two-Gaussian fits to the H$\alpha$
emission-line profiles in NGC~2992. The positions relative to the
radio nucleus are indicated at the top of each panel, the velocity
scale relative to systemic is at the bottom. }
\end{figure}

\begin{figure}
\caption{Distribution of the H$\alpha$ and continuum emission in
NGC~2992.  ($a$) Continuum emission at rest wavelength $H\alpha$,
($b$) total H$\alpha$ emission, ($c$) H$\alpha$ emission from the disk
component, ($d$) H$\alpha$ emission from the outflow component. The
position of the radio nucleus (``+'') and the extent of the dust lane
are indicated on each of these panels. North is at the top and east to
the left. Note the absence of a disk component in the east
quadrant. }
\end{figure}

\begin{figure}
\caption{Contour maps of the X-ray emission (left panel: ROSAT/HRI
data from Colbert et al. 1998) and 6-cm continuum emission (upper
right panel: VLA C-configuration data from Colbert et al. 1996a with
3$\arcsec$ uniform weighting; lower right panel: VLA A-configuration
data from Ulvestad \& Wilson 1984) superimposed on the total H$\alpha$
emission from NGC~2992 (grey-scale). A cross (``+'') in the main
figure indicates the position of the radio nucleus. The arrows in the
right panels mark the direction of the one-sided 90$\arcsec$ (13.5
kpc) radio extension along P.A. $\sim$ 100$\arcdeg$ -- 130$\arcdeg$,
i.e. close to the axis of the biconical outflow. Same orientation as
Fig. 2.}
\end{figure}

\begin{figure}
\caption{(color) Velocity fields of the disk ($a$) and outflow ($b$,
$c$) components. The velocity scale is shown at the bottom of the figure.
The line of nodes derived from the disk component is shown
superposed as a series of black dots in panel $a$. A cross (``+'')
indicates the position of the radio nucleus in each panel. A second,
high-velocity redshift component is seen in the SE cone and on the
northern tip of the NW cone. Same orientation as Fig. 2. }
\end{figure}

\begin{figure}
\caption{(color) Line widths (FWHM) of the disk ($a$) and outflow
($b$, $c$) components. These widths were corrected in quadrature for
the instrumental profile (FWHM$_i$ = 50 km s$^{-1}$). The velocity
scale is shown at the bottom of the figure. A cross (``+'') indicates
the position of the radio nucleus in each panel. Same orientation as
Fig. 2.}
\end{figure}

\begin{figure}
\caption{The filled circles show the rotation curve of NGC~2992
derived along the line of nodes (line of steepest gradient) of the
disk component. The filled triangles show the velocities along the
line of shallowest gradient. The dashed line corresponds to the
systemic velocity, 2,335 km s$^{-1}$, derived from model fitting of
the 2D velocity field (see Fig. 7). Note the asymmetry between the two
sides of the rotation curve.}
\end{figure}

\begin{figure}
\caption{(color) ($a$) Predicted velocity field for an axisymmetric
disk.  ($b$) Residual map (= observed H$\alpha$ velocity field $-$
model).  The parameters of the model are described in \S 4.2. A cross
(``+'') indicates the position of the radio nucleus.  This simple
axisymmetric model fits the data remarkably well. The largest positive
and negative residuals between the data and the model are seen $\sim$
10$\arcsec$ -- 20$\arcsec$ from the nucleus along P.A. $\approx$
40$\arcdeg$ -- 220$\arcdeg$, and are due in the most part to our use
of a symmetric rotation curve in the model.}
\end{figure}

\begin{figure}
\caption{Geometry of the idealized biconical outflow.  The angles
indicated on this figure and derived in the text are $i = 68\arcdeg,
\theta = 130\arcdeg, \alpha = 3\arcdeg, {\rm and}~\beta = 47\arcdeg$.
In the simplest form of the model, the gas flows along the surface of
the bicone and the outflow velocities are symmetric with respect to
the galaxy disk. The Fabry-Perot data do not agree in detail with this
simple picture. The line-emitting material appears to partially
fill-in the volume of the bicone in NGC~2992 and the kinematic
component on the near side of the NW cone is undetected. Asymmetry in
the outflow velocities appears to be present in this galaxy: $V_1$ =
25 -- 100 km~s$^{-1}$, $V_2$ = 40 -- 150 and $V_4$ = 25 -- 220
km~s$^{-1}$. There may also be a trend for decreasing outflow
velocities with increasing distance from the nucleus (more evident in the
$V_2$ and $V_4$ components).}
\end{figure}

\clearpage

\setcounter{figure}{0}
\begin{figure}
\caption{}
\end{figure}

\begin{figure}
\caption{}
\end{figure}

\begin{figure}
\caption{}
\end{figure}

\begin{figure}
\caption{}
\end{figure}

\begin{figure}
\caption{}
\end{figure}

\begin{figure}
\plotone{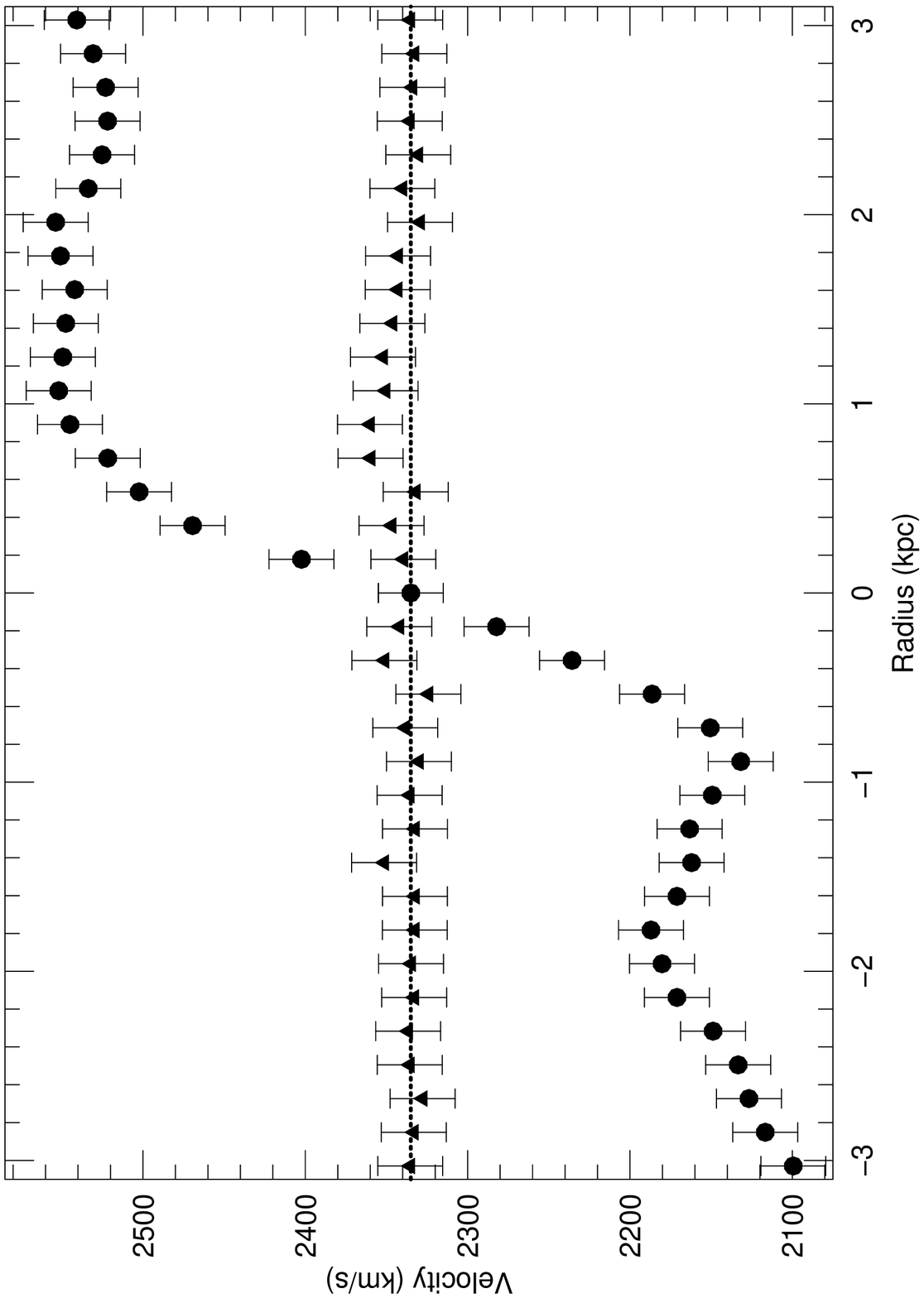}
\caption{}
\end{figure}

\begin{figure}
\caption{}
\end{figure}

\begin{figure}
\plotone{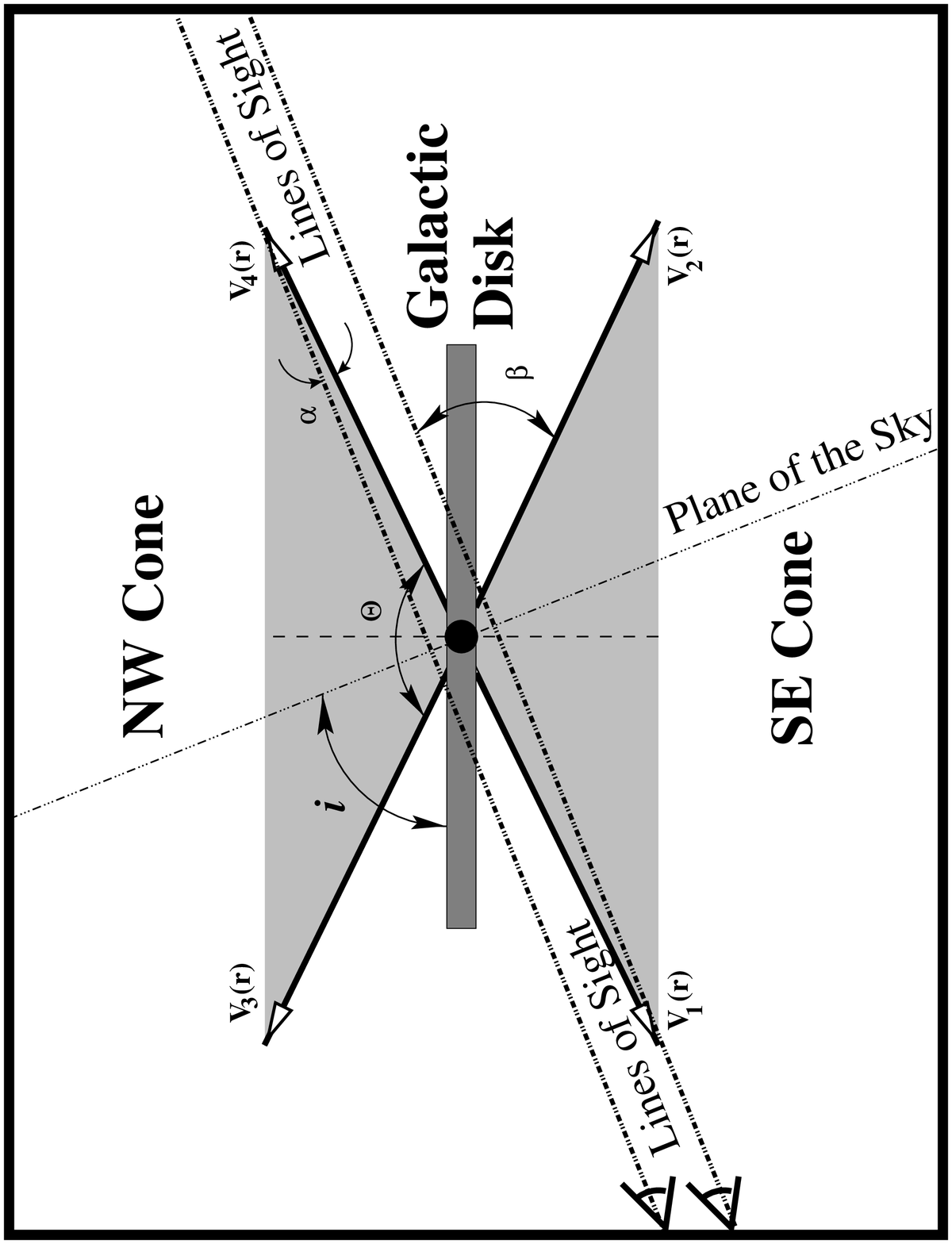}
\caption{}
\end{figure}

\clearpage


\clearpage

\begin{references}

\refpar
Allen, M. G. 1998, Ph. D. Thesis, Australian National University
\refpar
Allen, M. G., Dopita, M. A., Tsvetanov, Z. I., \& Sutherland, R. S. 1999, ApJ, 
511, 686
\refpar
Arp, H. C. 1966, Atlas of Peculiar Galaxies
\refpar
Axon, D. J., Marconi, A., Capetti, A., Maccetto, F. D., Schreier, E., \& Robinson, A. 1998, ApJ, 496, L75
\refpar
Balsara, D. S., \& Krolik, J. H. 1993, ApJ, 402, 109
\refpar
Baum, S. A., O'Dea, C. P., Dallacassa, D., de Bruyn, A. G., \& Pedlar, A. 
1993, ApJ, 419, 553
\refpar
Bicknell, G. V., Dopita, M. A., Tsvetanov, Z. I., \& Sutherland, R. S. 1998, ApJ, 495, 680
\refpar
Bland J. 1986, Ph.D. thesis, University of Sussex
\refpar
Bland, J., \& Tully, R. B. 1988, Nature, 334, 43
\refpar
Bland, J., \& Tully, R. B. 1989, AJ, 98, 723
\refpar
Bland-Hawthorn, J. 1995, in Trimensional Spectroscopic Methods in 
Astrophysics, IAU Coll. 149, ASP Conf. Ser. Vol. 71, eds. G. Comte and 
M. Marcelin, p. 72
\refpar
Bower, G., Wilson, A., \& Mulchaey, J. 1994, AJ, 107, 1686
\refpar
Bower, G., Wilson, A., Morse, J. A., Gelderman, R., Whittle, M., \& Mulchaey, J. 1995, ApJ, 454, 106
\refpar
Brandt, J. C. 1960, ApJ, 131, 293
\refpar
Bregman, J. N. 1978, ApJ, 224, 768
\refpar
Burbidge, E. M., Strittmatter, P. A., Smith, H. E., \& Spinrad, H. 1972, ApJ, 178, L43
\refpar
Burstein, D., \& Heiles, C. 1982, AJ, 87, 1165
\refpar
Capetti, A., Macchetto, F., Axon, D. J., Sparks, W. B., \& Boksenberg, A. 1995a, ApJ, 448, 600
\refpar
Capetti, A., Axon, D. J., Kukula, M., Macchetto, F., Pedlar, A., Sparks, W. B., \& Boksenberg, A. 1995b, ApJ, 454, L85
\refpar
Carollo, C. M., \& Danziger, I. J. 1994, MNRAS, 270, 523
\refpar
Cecil, G. 2000, Imaging the Universe in Three Dimensions, ASP Conf. Ser. Vol. 195, eds. W. van Breugel and J. Bland-Hawthorn, p. 263
\refpar
Cecil, G., Bland, J., \& Tully, R. B. 1990, ApJ, 329, 38
\refpar
Cecil, G., Veilleux, S., Bland-Hawthorn, J., \& Filippenko, A. V. 2000, ApJ, 
submitted
\refpar
Cecil, G., Wilson, A. S., \& Tully, R. B. 1992, ApJ, 390, 365
\refpar
Chambers, K. C., Miley, G. K., \& van Breugel, W. 1987, Nature, 329, 604
\refpar
Chapman, S. C., Morris, S. L., Alonso-Herrero, A., \& Falcke, H. 2000, MNRAS, 314, 263
\refpar
Chevalier, R. A., \& Clegg, A. W. 1985, Nature, 317, 44
\refpar
Colbert, E. J. M., Baum, S. A., Gallimore, J. F., O'Dea, C. P., \& Christensen, J. A. 1996a, ApJ, 467, 551
\refpar
Colbert, E. J. M., et al. 1996b, ApJS, 105, 75
\refpar
Colbert, E. J. M., Baum, S. A., O'Dea, C. P., \& Veilleux, S. 1998, ApJ, 496, 786
\refpar
Colina, L., Fricke, K. J., Kollatschny, W., \& Perryman, M. A. C. 1987, A\&A, 178, 51
\refpar
Condon, J. J., Condon, M. A., Gisler, G., \& Puschell, J. J. 1982, ApJ, 252, 102
\refpar
de Vaucouleurs, G., de Vaucouleurs, A., H. G. Jr. Corwin, Buta, R. J.,
Paturel, G., \& Fouqu\'e, P. 1991, Third Reference Catalogue of Bright
Galaxies (New York, Springer-Verlag)
\refpar
Dopita, M. A., \& Sutherland, R. S. 1996, ApJS, 102, 161
\refpar
Durret, F. 1990, A\&A, 229, 351
\refpar
Durret, F., \& Bergeron, J. 1987, A\&A, 173, 219
\refpar
Durret, F., \& Bergeron, J. 1988, A\&A Suppl., 75, 273
\refpar
Elmouttie, M., Haynes, R. F., Jones, K. L., Sadler, E. M., \& Ehle, M. 1998, MNRAS, 297, 1202
\refpar
Ferruit, P., Binette, L., Sutherland, R. S., \& P\'econtal, E. 1997a, A\&A, 322, 73
\refpar
Ferruit, P., P\'econtal, E., Wilson, A. S., \& Binette, L. 1997b, A\&A, 328, 493 
\refpar
Fiedler, R., \& Henriksen, R. N. 1984, ApJ, 281, 554
\refpar
Franx, M., \& Illingworth, G. 1990, ApJ, 359, L41
\refpar
Gallimore, J. F., Baum, S. A., \& O'Dea, C. P. 1996, ApJ, 464, 198
\refpar
Gallimore, J. F., Baum, S. A., O'Dea, C. P., Pedlar, A., Brinks, E. 1999, ApJ, 524, 684
\refpar
Goodrich, R. W., Veilleux, S., \& Hill, G. J. 1994, ApJ, 422, 521
\refpar
Haniff, C. A., Wilson, A. S., \& Ward, M. J. 1988, ApJ, 334, 104
\refpar
Heckman, T. M., Armus, L., \& Miley, G. K. 1990, ApJS, 74, 833
\refpar
Heckman, T. M., Miley, G. K., van Breugel, W. J. M., \& Butcher, H. R. 1981, ApJ, 247, 403
\refpar
Henriksen, R. N., Vall\'ee, J. P., \& Bridle, A. H. 1981, ApJ, 249, 40
\refpar
Hutchmeier, W. K. 1982, A\&A, 110, 121
\refpar
Hummel, E., van Gorkom, J. H., \& Kotanyi, C. G. 1983, ApJ, 267, L5
\refpar
Jablonka, P., Martin, P., \& Arimoto, N. 1996, AJ, 112, 1415
\refpar
J{\o}rgensen, I., Franx, M., \& Kjaergaard, P. 1996, MNRAS, 280, 167
\refpar
Keel, W. C. 1996, ApJS, 106, 27
\refpar
Krolik, J. H., \& Begelman, M. C. 1986, ApJ, 308, L55
\refpar
Kukula, M. J., Ghosh, T., Pedlar, A., Schilizzi, R. T. 1999, ApJ, 518, 117
\refpar
Larson, R. B., \& Dinerstein, H. L. 1975, PASP, 87, 911
\refpar
Lehnert, M. D.,  \& Heckman, T. M. 1996, ApJ, 462, 651
\refpar
MacLow, M.-M., \& McCray, R. 1988, ApJ, 324, 776
\refpar
MacLow, M.-M., McCray, R., \& Norman, M. L. 1989, ApJ, 337, 141
\refpar
Maiolino, R., Ruiz, M., Rieke, G. H., \& Papadopoulos, P. 1997, ApJ, 485, 552
\refpar
Marlowe, A. T., Heckman, T. M., Wyse, R. F. G., \& Schommer, R. 1995, ApJ, 438, 563
\refpar
M\'arquez, I., Boisson, C., Durret, F., \& Petitjean, P. 1998, A\&A, 333, 459
\refpar
Meurer, G. R., Freeman, K. C., Dopita, M. A., \& Cacciari, C. 1992, AJ, 103, 60
\refpar
Mineshige, S., Shibata, K., \& Shapiro, P. R. 1993, ApJ, 409, 663
\refpar
Nelson, C. H., \& Whittle, M. 1995, ApJS, 99, 67
\refpar
Osmer, P. S., Smith, M. G., \& Weedman, D. W. 1974, ApJ, 192, 279
\refpar 
Osterbrock, D. E. 1989, Astrophysics of Gaseous Nebulae and Active Galactic Nuclei, University Science Books
\refpar
Osterbrock, D. E., \& Martel, A. 1992, PASP, 101, 76
\refpar
Pahre, M. A., Djorgovski, S. G., \& de Carvalho, R. R. 1998, AJ, 116, 1591
\refpar
Pedlar, A., Dyson, J. E., \& Unger, S. W. 1985, MNRAS, 214, 463
\refpar
Pedlar, A., et al. 1993, MNRAS, 263, 471 
\refpar
Rix, H.-W., Carleton, N. P., Rieke, G., \& Rieke, M. 1990, ApJ, 363, 480
\refpar
Sadler, E. M., Slee, O. B., Reynolds, J. E., Roy, A. L. 1995, MNRAS, 276, 1373
\refpar
Sanders, D. B., \& Mirabel, I. F. 1985, ApJ, 298, L31
\refpar
Schlegel, D. J., Finkbeiner, D. P., \& Davis, M. 1998, ApJ, 500, 525
\refpar
Schiano, A. V. R. 1985, ApJ, 299, 24
\refpar
Shopbell, P. L., \& Bland-Hawthorn, J. 1998, ApJ, 493, 129
\refpar
Shopbell, P. L., et al. 2000, ApJ, submitted
\refpar
Shuder, J. M. 1980, ApJ, 240, 32
\refpar
Slavin, J. D., \& Cox, D. P. 1992, ApJ, 392, 131
\refpar
Smith, M. D., \& Norman, C. A. 1981, MNRAS, 194, 771
\refpar
Spinoglio, L., \& Malkan, M. A. 1989, ApJ, 342, 83
\refpar
Staveley-Smith, L., Bland, J., Axon, D. J., Davies, R. D., \& Sharples, R. M. 1990, ApJ, 364, 23
\refpar
Steffen, W., G\'omez, J. L., Williams, R. J. R., Raga, A. C., \& Pedlar, A. 1997, MNRAS, 286, 1032
\refpar
Strickland, D. K., \& Stevens, I. R. 2000, MNRAS, 314, 511
\refpar
Suchkov, A. A., Balsara, D. S., Heckman, T. M., \& Leitherer, C. 1994, ApJ, 430, 511
\refpar
Suchkov, A. A., Berman, V. G., Heckman, T. M., \& Balsara, D. S. 1996, ApJ, 463, 528
\refpar
Taylor, K., \& Atherton, P. D. 1980, MNRAS, 191, 675
\refpar
Taylor, D., Dyson, J. E., Axon, D. J., \& Pedlar, A. 1989, MNRAS, 240, 487
\refpar
Taylor, D., Dyson, J. E., \& Axon, D. J. 1992, MNRAS, 255, 351
\refpar
Tenorio-Tagle, G., \& Bodenheimer, P. 1988, ARAA, 26, 45
\refpar
Tomisaka, K. 1990, ApJ, 361, L5
\refpar
Tomisaka, K., \& Ikeuchi, S. 1988, ApJ, 330, 695
\refpar
Trager, S. C., Worthey, G., Faber, S. M., Burstein, D., \& Gonzalez, J. J. 1998, ApJS, 116, 1
\refpar
Ulvestad, J. S., \& Wilson, A. S. 1984, 285, 439
\refpar
Ulvestad, J. S., \& Wilson, A. S. 1989, 343, 659
\refpar
Unger, S. W., Pedlar, A., Axon, D. J., Whittle, M., Meurs, E. J. A., \& Ward, M. 1987, MNRAS, 228, 671
\refpar
Unger, S. W., Pedlar, A., Booler, R. V., Harrison, B. A. 1986, MNRAS, 219, 387
\refpar
Vader, P. 1986, ApJ, 305, 669
\refpar
Veilleux, S. 2000, Imaging the Universe in Three Dimensions, ASP Conf. Ser. Vol. 195, eds.  W. van Breugel and J. Bland-Hawthorn, p. 277
\refpar
Veilleux, S., \& Bland-Hawthorn, J. 1997, ApJ, 479, L105
\refpar
Veilleux, S., Bland-Hawthorn, J., \& Cecil, G. 1999, ApJ, 520, 111
\refpar
Veilleux, S., Cecil, G., Bland-Hawthorn, J., Tully, R. B., Filippenko,
A. V, \& Sargent, W. L. W. 1994, ApJ, 433, 48
\refpar
Veilleux, S., Goodrich, R. W., \& Hill, G. J. 1997, ApJ, 477, 631
\refpar
Veilleux, S., Tully, R. B., \& Bland-Hawthorn, J. 1993, AJ, 105, 1318
\refpar
V\'eron, P., Lindblad, P. O., Zuiderwijk, E. J., V\'eron, M.-P., \& Adam, G. 1980, A\&, 87, 245
\refpar
Ward, M., Penston, M. V., Blades, J. C., \& Turtle, A. J. 1980, MNRAS, 193, 563
\refpar
Ward, M., Wilson, A. S., Penston, M. V., Elvis, M., Maccacaro, T., \& Tritton, K. P. 1978, ApJ, 223, 788
\refpar
Weaver, K., et al. 1996, ApJ, 458, 160
\refpar
Wehrle, A. E., \& Morris, M. 1988, AJ, 95, 1689
\refpar
Wei\ss, A., Walter, F., Neininger, N., \& Klein, U. 1999, A\&A, 345, L23
\refpar
Whittle, M., et al. 1986, MNRAS, 222, 189
\refpar
Whittle, M., Pedlar, A., Meurs, E. J. A., Unger, S. W., Axon, D. J., \& Ward, M. J. 1988, ApJ, 326, 125
\refpar
Wilson, A. S. 1981, in Proc. Second ESO/ESA Workshop on Optical Jets
in Galaxies, ed. A. F. M. Moorwood \& K. Kj\"ar (Garching: ESO), 125
\refpar
Wilson, A. S., Elvis, M., Lawrence, A., \& Bland-Hawthorn, J. 1992, ApJ, 391, L75
\refpar
Wilson, A. S., \& Tsvetanov, Z. 1994, \aj, 107, 1227
\refpar
Zaritsky, D., Kennicutt, R. C. Jr, \& Huchra, J. P. 1994, ApJ, 420, 87

\end{references}
\end{document}